\PassOptionsToPackage{hyphens}{url}

\documentclass[]{interact}
\DeclareUnicodeCharacter{0301}{\'{}}  

\usepackage{microtype}

\usepackage{epstopdf}
\usepackage[caption=false]{subfig}

\usepackage[natbibapa,nodoi]{apacite}
\theoremstyle{plain}

\theoremstyle{definition}

\theoremstyle{remark}

\usepackage{comment}

\begin{document}


\title{Metamathematics of Algorithmic Composition}

\author{
\name{Michael Gogins\textsuperscript{a}\thanks{CONTACT Michael Gogins Email: michael.gogins@gmail.com}}
\affil{\textsuperscript{a}Irreducible Productions, 1576 Crescent Valley Road, Bovina Center, New York 13740, United States}
}

\maketitle

\begin{abstract}
I recount my journey towards a deeper understanding of the philosophical context, mathematical foundations, and computational complexity of algorithmic music composition. I discuss different types of algorithms, but my primary focus is on the fundamental limits and possibilities of algorithmic composition, by analogy with metalogic, metamathematics, and computability theory. I present implications for the practice and future of algorithmic composition. 
\end{abstract}

\section{Introduction}

In 1983, as a returning undergraduate in comparative religion at the University of Washington, I attended a lecture on fractals by Benoit Mandelbrot, discoverer of the eponymous set \citep{citeulike:580392, peitgen2004mandelbrot}. Briefly, given the quadratic recurrence $z_{n+1} = z_n^2 + c$, the Mandelbrot set consists of all points $c$ in the complex plane for which $z$, starting at 0, does not approach infinity as the equation is iterated. Then, for each point $c$ in the Mandelbrot set there is a connected Julia set, consisting of all points $z$ for which $z_n$ does not approach infinity as the equation is iterated. Mandelbrot's slides showed how a plot of the neighborhood near some $c$ in the Mandelbrot set closely resembles the plot of the corresponding Julia set (\citet{lei1990similarity}; recently proved by \citet{kawahira2018julia}). In short, the Mandelbrot set is a \emph{parametric map} of all connected Julia sets. There is now an extensive and growing literature on the Mandelbrot set and Julia sets. 

I was already interested in computer music, especially algorithmic composition based on fractals. During the lecture it occurred to me that if I zoomed into a plot of the Mandelbrot set, searching for interesting-looking regions, I could plot the Julia set for a point in that region, and then somehow translate that Julia set into a musical score \citep{obsessed}. I have worked out several variations of this idea.

\subsection{Mandelbrot Set/Julia Set}

The composer explores the Mandelbrot set. When a region seems interesting, the composer selects a point in it, and the corresponding Julia set is first generated, and then translated to a musical score. The 2-dimensional plot of the Julia set is mapped onto a 3-dimensional piano-roll score: the $x$ axis is time, the $y$ axis is pitch, and the $z$ axis is the choice of instrument, also indicated by color.

\subsection{Parametric Map of Iterated Function Systems}

An iterated function system (IFS) is a Hutchinson operator, a set of contractive affine transformations, each of which is applied to transform an initial set; the union of these transformations becomes the initial set for the next iteration. Iterating the operator upon \emph{any} initial set takes that set to the \emph{same} fixed point, which is a fractal \citep{barnsley1985iterated, barnsley1993}. The Collage Theorem proves that an IFS can approximate any set as closely as desired \citep{barnsley1989fractal, barnsley1993}. Therefore this method is \emph{compositionally universal} \citep{obsessed, gogins2023scoregraphs}: it can generate, as closely as one likes, any finite score. 

It is also possible to generate a parametric map of any subset of IFSs, but as Hutchinson operators have more than two parameters, generating a parametric map (an analogue of the plot of the Mandelbrot set) for dozens or hundreds of parameters requires the use of a Hilbert index, which maps points in an $N$-dimensional space, such as a plane, cube, or hypercube, to a 1- or 2-dimensional sequence of numbers \citep{patrick1968mapping}. 

There is one dimension for each IFS parameter. The index is constructed so that neighboring points in the $N$-dimensional space usually have neighboring entries in the 1- or 2-dimensional index. The key idea is to recursively subdivide the $N$-dimensional space into smaller planes, cubes, or hypercubes, called cells. Each cell is assigned a unique index based on its position within the overall space. Subdivision continues recursively until a desired level of detail is reached. To determine the Hilbert index of a specific point in the $N$-dimensional space, start with the largest cell containing the point, level $j = 0$. Then, working in arithmetic to the base $N$, subdivide that cell into $N$ sub-cells for level 1, and select the sub-cell containing the point. If it is the $kth$ sub-cell at level $j$, then add $((k + 1)/N)^{-j}$ to the index. Repeat this process recursively until the smallest cell containing the point is reached. The index of that cell is  the Hilbert index for the point. Hilbert indices work because all metric spaces in $\mathbb{R}^n$ have the same cardinality; hence there is always a one-to-one mapping between points in any $N$-dimensional space and points on the line or, more usefully, the plane. The plot of the IFS is translated to a score in much the same way as for the plot of a Julia set.

I experimented with both methods, doing parametric composition by zooming in on interesting regions of the map, generating and rendering scores, exploring points near scores that seemed promising, and iterating this process. However, I found that scores generated directly from Julia sets had too much of a sameness. And producing a parametric map of more than just a few points for IFSs simply took way too much time. In other words, computing parametric maps for IFSs is \emph{computationally intractable}. 

As I pursued algorithmic composition, I found that intractability is not merely a practical problem, but has profound mathematical and philosophical roots, beginning with Pythagoras and continuing on through the hierarchy of complexity classes in theoretical computer science. I found only a few discussions of these issues of complexity and intractability in the computer music literature.

\section{Philosophical Context}

Music has since Pythagoras \citep{sep-pythagoras, huffman2014history} been understood by some as an intellectual paradigm and to reveal, through harmony both numerical and audible, the structure of reality. For this reason music was a central part of the \emph{quadrivium}, the standard curriculum of liberal arts in Western higher education from late Antiquity through the Renaissance.

The project of understanding reality through number advanced from Pythagoras, through Leibniz' 
hope for a \emph{characteristica universalis}, a symbolic language that could express all rational thought \citep{davis2018universal}, to the logicism of Russell \citep{sep-logicism}, Hilbert \citep{sep-hilbert-program}, and others, that sought to derive all mathematical truths from formal logic. In one of the major achievements of all philosophy, Kurt Gödel \citep{godel1986} demonstrated in his incompleteness theorems that logicism cannot be fully implemented, because there exist true statements of logic that cannot  be proved. Five years later, Alan Turing proved the Halting Theorem: it is impossible for any computer program to decide whether another, arbitrary computer program will halt. Researchers are still exploring the consequences of these theorems. Some major results:

\begin{itemize}
\item The elucidation of a provable hierarchy of complexity classes for problems that are solvable (or not) by computer \citep{arora2009computational}.
\item Proofs that there are abstract (so far completely unphysical) computers that are \emph{not} incomplete; these are called super-Turing and depend, one way or another, on doing arithmetic with real numbers (almost all real numbers are uncomputable, because their decimal representations never terminate) \citep{ord2006many}.
\item Proofs that the relationship between energy and information imposes hard \emph{physical} limits on the amount of computation that can be done in the physical universe \citep{aaronson2005npcomplete, sep-computation-physicalsystems}.
\end{itemize}

 \subsection{Complexity Classes} \label{sec:complexity}
 
 The complexity classes are based on the capabilities and limitations of Turing machines. A TM is an abstract, idealized computer with infinite memory that can run for an infinite number of steps. It may halt after a finite number of steps and produce a result, or never halt. A \emph{universal} TM can emulate any other TM. The widely accepted Church-Turing thesis holds that \emph{anything} computable by a definite, step by step procedure is computable by a universal TM \citep{sep-church-turing}. Contemporary digital computers are universal TMs --- or would be, if they had infinite memory and infinite time to run.
 
 For our purposes, the important complexity classes are:
 
 \begin{description}
\item[$\mathsf{ST}$ (super-Turing)] Problems that have \emph{mathematical} solutions but are not solvable by any TM. These include problems that are neither recursively enumerable nor recursively computable. They lie beyond the scope of classical computation.
  
\item[$\mathsf{RE}$ (recursively enumerable)] Also called \emph{semi-decidable}. A TM can recognize solutions: if an input belongs to the language (\emph{i.e.}, the answer is ``yes''), the machine will eventually accept it. However, if the answer is ``no,'' the machine may run forever. Thus, non-membership is not decidable.

\item[$\mathsf{R}$ (recursively computable)] Also called \emph{decidable}. Problems for which a TM can decide in finite time whether a given input belongs to the language. These problems can be both accepted and rejected by some algorithm.

\begin{description}
\item[$\mathsf{P}$ (deterministic polynomial time)] Problems that can be solved by a deterministic TM in polynomial time. These are considered ``efficiently solvable.''

\item[$\mathsf{NP}$ (non-deterministic polynomial time)] Problems for which a candidate solution can be \emph{verified} in polynomial time by a deterministic TM. It is unknown whether every problem in $\mathsf{NP}$ can also be \emph{solved} in polynomial time (i.e., whether $\mathsf{P} = \mathsf{NP}$).

\item[$\mathsf{NP}$-complete] The hardest problems in $\mathsf{NP}$; if any one of them can be solved in polynomial time, then all $\mathsf{NP}$ problems can. Each $\mathsf{NP}$-complete problem is both in $\mathsf{NP}$ and $\mathsf{NP}$-hard.

\item[$\mathsf{PSPACE}$ (polynomial space)] Problems solvable by a TM using polynomial \emph{memory}, regardless of how much time it takes. This class includes all of $\mathsf{P}$ and $\mathsf{NP}$, but may also contain problems harder than $\mathsf{NP}$.

\end{description}
\end{description}

\noindent The complexity classes mirror central issues in philosophy and science. It is a primary open question of philosophy whether Nature herself is super-Turing. If so, then human thought, including musical composition, might as part of Nature also be super-Turing. If not, then human thought is at most $\mathsf{RE}$ and could be emulated as closely as one likes by a TM. 

Yet in either case, scientific theories are $\mathsf{RE}$ or less, because it must be possible to compare the predictions of the theory to observations of Nature that are finite in both number and precision. Nobody can see how Nature or human thought being super-Turing could ever be \emph{proved}, so it is often held that any physical process can be effectively simulated by a TM; this is the \emph{physical} Church-Turing thesis \citep{aaronson2005npcomplete, sep-church-turing}. Compositional algorithms that halt are $\mathsf{NP}$ or less.

Whether or not $\mathsf{NP}$ is contained in $\mathsf{P}$ is one of the most important open questions in science. Most mathematicians and scientists believe, for overlapping reasons, that $\mathsf{P} \neq \mathsf{NP}$. If $\mathsf{P} = \mathsf{NP}$  can be proved, it becomes possible in principle to automatically solve all problems of a size that human beings can grasp. But if $\mathsf{P} \ne \mathsf{NP}$ can be proved, we will know that some problems are too complex for computers to solve.
 
\subsection{Algorithmic Composition in Context}

We can now return to the analysis of algorithmic composition with more understanding. Some compositional algorithms use strictly \emph{finitary} methods, and others use implicitly \emph{infinitary} methods. Both finitary and infinitary algorithms can exhibit \emph{computational irreducibility} in the sense of Wolfram \citep{wolfram1985undecidability}. In order to know what any non-trivial algorithm actually does, one must actually run the algorithm.

Examples of finitary algorithms include automatic counterpoint generation based on Renaissance rules \citep{schottstaedt1989automatic, taube1991common}, generating scores using finite iterations of Lindenmayer systems \citep{mccormack1991lsystems}, and so on. Gwee formalized algorithms for constructing counterpoint given a \emph{cantus firmus}, constructed decision problems for those algorithms, and proved they are in $\mathsf{PSPACE}$ \citep{gwee2002complexity, gwee2013music}. This implies that rule- or constraint-based compositional problems are as hard as \emph{any} recursively computable problem.

Examples of infinitary algorithms include stochastic music computed with real-valued random variables \citep{xenakis2001formalized}, music computed using real-valued dynamical systems \citep{voss1975noise, gardner1978fractal, beyls1991chaos}, and my Mandelbrot set/Julia set and IFS ideas. 

Braverman proved that hyperbolic Julia sets are computable, and in $\mathsf{P}$ (measure 1); most parabolic Julia sets are computable, but in $\mathsf{NP}$ or worse (measure 0); and some Julia sets with Siegel discs are neither hyperbolic nor parabolic, but \emph{uncomputable} (measure 0) \citep{braverman2006non, braverman2008computability, braverman2009computability}.
\begin{description}
\item[$\mathsf{ST}$]  Some Julia sets with Siegel discs might be musically interesting in the abstract, but they are uncomputable. Picking Julia set parameters at random won't find any of these.
\item[$\mathsf{RE}$] The Mandelbrot set, properly speaking, is not \emph{recursively computable}, \emph{i.e.}\ not Turing computable \citep{blum1993godel}. Plots we make of the set are approximations. Hertling showed that although the Mandelbrot set is not recursively computable, it is nevertheless \emph{recursively enumerable} \citep{Hertling2005-HERITM-3}; given enough time, one can approximate the actual set as closely as one likes. 
\item[$\mathsf{NP}$] If some parabolic Julia sets are musically interesting, then computing a parametric map that is complete enough to be useful for composers might remain forever computationally intractable. For, as Julia set parameters are chosen closer and closer to a Siegel point, the Julia sets get harder and harder to compute, and the appearance of computer generated plots or scores fluctuates wildly. It is not clear whether $\mathsf{NP}$ includes any large number of musically interesting compositions.
\item[$\mathsf{P}$] But if $\mathsf{P} = \mathsf{NP}$ then in principle it is possible to produce a useful parametric map in polynomial time: no sets with Siegel discs or Cremer points, but all hyperbolic points and as close as we like to almost all parabolic points. In that case, parametric composition might open up a vast new world for composers.
\end{description}

Julia sets are not the only attractors of dynamical systems that can be $\mathsf{ST}$. Other types of dynamical systems that can have $\mathsf{ST}$ attractors include shift maps \citep{moore1991generalized}, piecewise affine maps \citep{Bazille_2018}, and continuous time recurrent neural networks \citep{10531294}. IFSs are recursively enumerable, but certain questions concerning their attractors are uncomputable \citep{dube1993undecidable}. The literature is scattered. The following are true for all these systems:

\begin{description}
\item[Universality] The degree to which compositional algorithms are effectively universal depends on how hard it is to compute attractors for parameter points that approach a parameter point having an computable attractor. The harder that is, the larger the intractable areas near the uncomputable points become, and the fewer pieces can be computed.
\item[Mappability] Only the ability to correlate numerical parameters for attractors with musical features of those attractors enables parametric composition. The degree of mappability depends on the computational complexity of attractors, especially those generated from parameters near the uncomputable points.
\item[Tractability] If $\mathsf{P} = \mathsf{NP}$, then a high-resolution parametric map of considerable subsets of compositions would display symmetries and patterns, such as those found in the Mandelbrot set, that could assist in parametric composition. It would afford a God's-eye view of (almost) all possible structure in music, and partly overcome the irreducibility of compositional algorithms. In other words, $\mathsf{P} = \mathsf{NP}$ could imply that algorithmic composition can be made \emph{intelligible}.

But if $\mathsf{P} \ne \mathsf{NP}$, parametric maps must exclude not only points with uncomputable attractors, but also larger neighborhoods of these points. This would to some extent fracture and obstruct the intelligibility of parametric composition.
\end{description}

The reader may well be wondering why I am talking about uncomputability when algorithmic compositions are normally \emph{computed}. The reason is, composers normally don't think about finitary or infinitary methods, or about computability or uncomputability. Indeed, composers often seem to simply assume continuous mathematics using real numbers. But all implicitly infinitary ideas must of course be implemented, using finitary methods, on a physical computer. 

We should always remember: things may be mathematically true, yet not computable; other things may be computable in the abstract, yet not computable in our physical world; and still other things may be physically computable, yet only given resources far exceeding ours. 

Beneath this issue is the older issue of philosophical realism (\emph{e.g.}\ considering even uncomputable Julia sets to be real in the Platonic sense, which ultimately derives from Pythagoreanism) versus pragmatism or nominalism (considering uncomputable Julia sets a convenient fiction). With respect to music, realism implies that a composer's idea is based on their grasp of an abstract yet real musical object. With respect to algorithmic composition, realism implies that pieces based on uncomputable attractors are nevertheless real musical objects, and neglecting them is ignorant, while pragmatism and nominalism say ``Just shut up and run the algorithm, the only pieces that matter are those we can compute.''

If neither can be proved, we will simply never know.

Another significant aspect of parametric composition in particular, and of algorithmic composition in general, is computational irreducibility. Almost every connected Julia set (measure 1) is the chaotic attractor of its generating recurrence. Therefore, the orbit of the Julia equation is computationally irreducible \citep{zwirn2015computational}. The orbit of the equation cannot be determined by examining the equation, and cannot except in trivial cases be determined even by mathematically analyzing the equation. In order to know the orbit, it is necessary to actually run a program that computes the orbit. Even then, we can only obtain an approximation. What is true of Julia sets is true of IFSs and, as many researchers believe, of many other algorithms.

In sum, uncomputability demonstrates fundamental limitations of compositional algorithms, and irreducibility demonstrates the only partial intelligibility of tractable algorithms.

\section{Methods of Algorithmic Composition}

Before further exploring the mathematical foundations of algorithmic composition, I will provide more background on algorithmic composition software.

I define \emph{algorithmic composition} as the use of computer software to write musical compositions that would be difficult or impossible to write without software. It does not include the use of notation software to write down music that one imagines (as that can be done with paper and pencil), nor the use of audio workstation software to record and overdub music that one improvises (as that can be done with a tape recorder). Algorithmic composition consists of all compositional techniques \emph{idiomatic} to the computer. There are many such \citep{fernandez2013ai, arizanet}. A recent summary is \citep{mclean2018oxford}. There is an obvious overlap with a more generic notion of \emph{process music} or \emph{generative music}, including Mozart's musical dice game \citep{humdrumdice}, the minimalism of Steve Reich \citep{reichprocess, 10.2307/832600} and Philip Glass \citep{potter2002four, glass2015words}, and the generative work of Brian Eno \citep{eno1996generative}. The commonality between algorithmic composition and process music is precisely the simplicity and clarity of the means, versus the complexity and unpredictability of the results; in other words, yet again, irreducibility. 

Irreducibility is a spectrum, not a binary choice. The minimum of irreducibility occurs when the composer simply writes down what they hear in their imagination. The maximum occurs when the composition is generated at random, so that there is no way for the composer to predict, better than chance, any particular note or sequence of notes; but even then, there is a degree of musical intelligibility in that the texture of one random variable (\emph{e.g.}\ white noise) sounds different from the texture of another random variable (\emph{e.g.}\ brown noise). In the middle of the spectrum is an area where the composer has \emph{some} insight into the kind of music that will be generated, even though details cannot be predicted. And this is the most interesting and most useful degree of computational irreducibility. 

Hiller and Isaacson's \emph{Illiac Suite} \citep{illiacsuite} is the first piece that is unambiguously computer music. The suite is an algorithmic composition assembled using a toolkit of stochastic note generators and music transformation functions, as detailed in \emph{Experimental Music} \citep{hiller}. This can be called the \emph{toolkit approach} to algorithmic composition. The composer programs a chain of generators, including random variables, and transformations, including serial transformations, to produce a chunk of music. The chunks can then be edited by hand. Multiple chunks can be assembled into a composition by hand. The toolkit approach lives on in contemporary software systems like Open Music \citep{OpenMusic}, Common Music \citep{CommonMusic, musx}, and many others. This is to date the most widely used method of algorithmic composition. 

\subsection{Algorithms in the Toolkit}

The toolkit approach includes generators and transformations from many sources:

\begin{description}
\item[Traditional Music Theory] Fugue and other canonical forms, scales, transpositions and rescalings of pitch and time.
\item[``Atonal Theory'' or ``Set Theory''] Transpose, invert, and reverse, and other group actions on tone rows or other pitch collections \citep{rahn1991basic}
\item[Neo-Riemannian Theory] Mathematizes voice-leading and discovers groups acting on notes, scales, and chords, together with their symmetries \citep{tymoczko2006geometry, tymoczko2011geometry, generalizedvoiceleadingspaces}.
\item[Random Variables] Borrowed from mathematics and used to generate series of musical properties such as pitches.
\item[Dynamical Systems and Cellular Automata] Borrowed from mathematics and used to generate series of musical properties such as pitches \citep{Miranda1993}.
\item[Fractals] Borrowed from mathematics and used to generate some or all aspects of musical scores \citep{miranda2001composing, madden2007fractals}. Some composers, including myself, prefer to use an algorithm, such as a Lindenmayer system \citep{algorithmicbeautyofplants, prusinkiewicz1986sgs,  fractalmusicwithstringrewritinggrammars} or IFS \citep{barnsley1993, ifsmusic} that will generate an entire piece based on fractals or other mathematical methods, needing no further editing or assembling. 
\item[Evolutionary Computing] Applying evolutionary computing to any of the above \citep{miranda2007evolutionary}.
\end{description}

\subsection{Intrinsic vs.\ Extrinsic Algorithms}

Most algorithms in the toolkit are computationally irreducible. However, there is a critical difference between applying, \emph{e.g.}, a sequence of transformations from the General Contextual Group \citep{fiore2005gcg}, even if the transformations are selected at random, versus simply sampling a random variable to sequentially scramble or transpose a chord. I define as \emph{intrinsic} all algorithms that act on primitive elements of music and generate musically acceptable transformations of those primitives, and \emph{extrinsic} all algorithms that are based on non-musical primitives, use transformations of the primitives that are not necessarily musically acceptable, or require mapping the output of the algorithm to musical scores.

In other words, while both procedures are computationally irreducible, any sequence of GCG actions will generate a much higher proportion of musically acceptable results than a sequence of random chords. That is because the GCG models some aspects of common musical practice. Other such syntaxes include abstracting the pragmatic rules of voice-leading, abstracting the network of chord progressions by scale degree in functional harmony, and so on.

Note well, a fractal or other mathematically derived algorithm may be modified to be based on, or to transform, musical primitives.

\subsection{Live Coding}

A more recent style of algorithmic composition known as \emph{live coding} extends the toolkit approach. A live coding system consists of a library of routines that are assembled into a music-generating graph during a live performance by interpreting real-time commands in a domain-specific language. Such systems have tools and commands for both high-level representations of music (notes, loops, chords, scales, musical transformations, \emph{etc.}) and sound synthesis (oscillators, envelope generators, filters, \emph{etc.}). An overview of the field can be gleaned from the TOPLAP web site \citep{toplap} and the \emph{Oxford Handbook of Algorithmic Composition} \citep{mclean2018oxford}.

\subsection{Artificial Intelligence}

Recently it has become possible to compose music using large language models (LLMs) like ChaptGPT. This can be called the \emph{artificial intelligence} (AI) approach to computer music. I discuss only LLMs as they are currently the most influential method of AI. Briefly, the approach is based on biology --- simulating at a high level of abstraction the behavior of nerve cells. A neural network is built up from layers of simulated neurons that connect with each other; the connections have tunable weights that control the output behavior of neurons in one layer given inputs from connected neurons in other layers. First the data, a large corpus of texts, is syntactically broken up into tokens. ``Attention heads'' are trained to learn weights indicating how tokens within the entire data set are related syntactically and semantically; each head looks at a different aspect of the data. The trained heads are then used to transform the input (the prompt) \emph{plus} the current output (hence the term ``\emph{self}-attention'') into a new input, from which the next token of output is predicted \citep{vaswani2017attention}. This attention mechanism, and other heuristics, have been found to greatly increase the power of the neural networks. In particular, the attention mechanism makes it possible to train the network on a huge body of data more efficiently and without much human intervention. For more detail, see \citep{zhang2023complete} and OpenAI's paper on their current LLM architecture \citep{openai2023gpt4}. For working examples of how LLMs can be used to compose music, see Jukebox \citep{openai2023jukebox}, Gonsalves \citep{aitunes}, and Ocampo \emph{et al.} \citep{ocampo2023using}.

Although it is early days for AI, there are some reviews and critical studies of the capabilities and limitations of LLMs. From the skeptical side, see \citep{dale2021gpt}. For an amusing series of dialogues between all sides, see \citep{aaronson2023should}. This experience makes it possible to identify some important limitations of AI:

\begin{description}
\item[Computational opacity] All agree that LLMs can generate amazing, even spooky, results without anyone understanding much about what is going on in the neural network. We have a perfect understanding of each component of the LLM, because these are actually quite simple, but we have no real idea how an LLM like ChatGPT can conduct a fact-filled conversation with one in perfect English. The details are scattered through hundreds of billions of neural network weights in the LLM. Computational opacity goes far beyond computational irreducibility --- we have taken one irreducible program (the untrained LLM) and used it to build another irreducible program (the trained LLM)! Even though an untrained LLM is computationally irreducible, we still have a perfect understanding of how it actually works step by step; but it seems very likely that we will not, at least in practice, ever obtain even a partial understanding of how the \emph{trained} LLM actually works.
\item [Hallucination] LLMs too often generate factually incorrect responses to prompts. It is a reminder that the software has no way to compare its responses with the real world. I suspect hallucinations arise from human mistakes, conflicting goals, and outright lies in the training data. Ways of dealing with hallucinations are being investigated; for one approach, see \citep{christiano2017deep}.
\item [Unoriginality] LLMs generate responses to prompts based on high-dimensional correlations that the LLMs have automatically discovered in the training data --- data \emph{we} have provided. This is a self-referential situation. When we converse with ChatGPT, we are looking at ourselves in a fun-house mirror.
\end{description}

\section{Artistic Results and Procedures}

To date, not many algorithmically composed pieces have become popular or influential, even among aficionados of art music and experimental music. A few paradigmatic pieces are Iannis Xenakis' \emph{La Légende D'Eer} \citep{Solr-8143160} and \emph{Gendy 3} \citep{gendy3},  Charles Dodge's \emph{Viola Elegy} \citep{violaelegy}, and Brian Eno's generative works \citep{eno1996generative, enochilvers}. I have my own idiosyncratic list of different algorithmic composition systems, with my own choice of representative pieces \citep{rant}.

The actual procedures followed by composers for algorithmic composition vary by genre, by composer, and by software used. IRCAM has published a series of books with chapters by composers on their use of OpenMusic \citep{omcomposersbook, agon2006om, agon2008om, agon2016om}. These are very useful. Profiles of composers in \emph{Computer Music Journal} are also useful. Here I outline the general procedure that I myself follow.

I start with some kind of mathematical system that can be used to generate a set of musical notes, a score. The system needs to generate a complex structure that can be changed by varying a relatively small number of numerical parameters or commands. It’s often useful to select a recursive algorithm that, as the number of iterations approaches infinity, approaches a fixed point that is a fractal. 

Such generative algorithms generally reflect processes in Nature that produce fractal-like forms, such as patterns on seashells or the branching of plants. I have used chaotic dynamical systems, Lindenmayer systems, IFSs, and other systems. And I have increasingly incorporated intrinsically musical primitives and transformations into, or along with, my algorithms (\citet{Gogins2020}, code at \citet{GoginsPoustiniaCode2025}).

Generally speaking, how to set the parameters in order to obtain a desired result is more or less opaque. This is well-known as the \emph{inverse problem} \citep{graham2021applying, tu2023learning}. But actually this is another form of computational irreducibility, meaning in this case it is not intuitive how to infer the structure of an algorithm even after closely inspecting its results. 

On the one hand this is a fault of the method; but on the other hand, and even more so, it is a virtue. In this way, and \emph{only} in this way, can we generate scores that we would not otherwise be able to imagine. This, of course, is another fundamental motivation for pursuing algorithmic composition. And it's the most important motivation. \emph{This kind of algorithmic composition actually amplifies the musical imagination} (see also \citet{edwards2011algorithmic}).

Now the question arises, how can such opaque methods be used to compose good music? It is difficult but by no means impossible.

The parameters generally have a global effect on the final structure. For example, an IFS consists of a number of affine transformations that are repeatedly applied to a set. Changing just one element of one transformation matrix can easily change every note in the generated score.

So, I pick one parameter and change its value, then listen to the result. I change the value again, and listen to the second result. Then, I choose the value that I prefer. I make more changes and listen again. Eventually I will find a value for that parameter that is more or less optimal – a ``sweet spot'' in the parameter space.

Then I pick another parameter and change its value in the same way, until I have a second sweet spot. During this process, the effect of the first parameter will probably change, so I go back to the first parameter and search for its sweet spot again.

This can be repeated for any number of parameters, but it is a tedious process and does not make sense for more than a few parameters.

This procedure amounts to a sort of binary search through a set of possible parameter values so vast – indeed infinite – that a linear search is simply out of the question. But a binary search is far more efficient than a linear search. Furthermore, finding two or three ``sweet spots'' in a small set of controlling parameters – each of which has global effects on the entire score – can produce a surprisingly large improvement in the musicality of the result.

I see here an analogy with the way in which LLMs work. There are repeated searches in a parameter space equipped with a fitness function (in LLMs, the loss function) at increasing levels of refinement (as with gradient descent).

Few if any algorithmic composers simply ``hear music in their heads'' and write software to render it. Most fool around producing various experimental chunks of music, refine them more or less as I have described, and assemble some into a finished composition.

\section{Conclusion}

Before I present my conclusions, I will summarize what I have learned about the mathematical foundations:

\begin{description}
\item[Uncomputability] The set of musical compositions that are possible \emph{in the abstract} is \emph{recursively uncomputable}.
\item[Universality] In spite of the uncomputability of the set of all compositions, any \emph{physically possible} composition is \emph{recursively enumerable} or less.
\item[Irreducibility] Most compositional algorithms are \emph{computationally irreducible}.
\item[Intrinsic vs.\ Extrinsic] Algorithms based on musical primitives and transformations produce a larger proportion of musically acceptable results.
\item[Opacity] Compositional algorithms based on AI are not only computationally irreducible, but also computationally opaque in that we have essentially no insight into the the steps followed by the LLM.
\item[Mappability] Compositional algorithms are nevertheless always mappable. This ultimately is because sets of compositions can be ordered in some way by \emph{musical} criteria, while their corresponding generating parameters can be ordered lexically or numerically.
\item[Intractability] Producing a useful parametric map of some subset of compositions is compute-intensive; applying complex rules to optimize a score can be in $\mathsf{PSPACE}$.
\item[Hallucination] LLMs that are supposed to provide true or useful outputs sometimes just make stuff up. Hence material generated by an LLM in response to a prompt cannot be trusted. A person, indeed an expert, must evaluate the response. It is by no means clear at this time whether an expert equipped with a LLM is more productive for creative work than that same expert without the LLM.
\item[Unoriginality] LLMs work by discovering high-dimensional correlations in large bodies of training data. LLMs can select, summarize, and vary --- but cannot generate a response that is not correlated with the training data. In other words, there is a limit to their originality. However, it is by no means clear at this time whether that limit is well below, or well above, the creativity of experts in the field from which the training data was drawn.
\end{description}

I will now put forward some conjectures based on these foundations.

\subsection{Limitations}

At this time and for the foreseeable future, no form of AI is conscious or has its own goals; AI also is subject to hallucination and unoriginality. Therefore, for the foreseeable future, human composers must and will play a irreplaceable role in algorithmic composition. This involves selecting a set of possible compositions to study, evaluating the musical quality of each composition in the subset, and varying the parameters or prompts to improve  the pieces.

Incomputability, intractability, opacity, and irreducibility set objective limits on how much understanding composers can gain into the working of their algorithms and of the music generated by them. This is both a limitation and an advantage. In practice, it is not possible to determine in advance just where those limits lie.

\subsection{Prospects}

Computer power will continue to increase. This will most likely make algorithmic composition both more productive and more important.

There is a similarity between a composer's experience with a toolkit of algorithms, the transformation of prompts into responses by an LLM, and exploring a parametric map for a fractal compositional algorithm. In all cases, starting with an initial trial, a final composition is approached by a descending, zigzag search through a multi-dimensional space representing musical possibilities of differing value, until the search comes to rest in some local optimum. This search can be greatly speeded up by using intrinsically musical algorithms that generate a larger proportion of musically acceptable results. For example, rather than representing scores as notes on piano rolls, \emph{e.g.}\ planes or cubes, one can represent scores as more or less fleeting chord progressions in chord spaces \citep{gogins2006score, gogins2023scoregraphs}.

Every method that speeds up searching makes algorithmic composition more productive. In particular, the growth of live coding demonstrates that the toolkit approach to algorithmic composition has a future. The underlying reason is that live coding, due to concise commands and immediately audible feedback, provides faster searching. Spending time doing live coding also increases the composer's insight into the tools.

\subsection{In Sum}

The major approaches to algorithmic composition --- trial and error with a toolkit of algorithms, live coding, exploration of fractals, and AI --- share the fundamental business of zigzagging down a slope in a multi-dimensional landscape of evaluations to rest in a local optimum. This result is proved by the simple fact that the generated music and/or the parameters used to generate it can be ordered. The dimensionality of the parameter space is secondary, as it can be reduced to one or two dimensions by means of a Hilbert index. Note that searching for solutions or optimizations in many domains is known to be $\mathsf{NP}$ or more.

Future developments in AI may have a significant impact on algorithmic composition. For example, AI has been applied to solving the inverse problem for discovering the parameters of fractal algorithms \citep{tu2023learning}. It might then be possible to represent an existing score, or scores, as fractal parameters and then work with these parameters to vary or interpolate between such pieces. This does not overcome computational irreducibility, as it substitutes the opacity of AI for the irreducibility of the inverse problem, yet it still might be very useful.

 Algorithmic compositions based on current LLMs are easier to produce, but seldom musically original; while algorithmic compositions based on toolkits, live coding, or fractals can be original, but are much more difficult to produce.

Usually the only way to penetrate the fog of incomputability, intractability, opacity, and irreducibility of the algorithms is to explore the geometrical order in a subset of compositions. This can be done either by trial and error, or by literally plotting a map of the subset. One might say that with trial and error one plots a sparse map of fully defined features in a territory, and with a parametric map one explores a densely mapped territory with partially defined features.

Progress in algorithmic composition seems likely to depend on:

\begin{itemize}
\item The increasing power of computer hardware, required to enable everything else.
\item Speeding up the composer's workflow, whether in parametric composition, algorithmic composition toolkits (\emph{e.g.}\ \citet{bellingham2019toward}), live coding, or AI.
\item Defining more musically compact and intelligible spaces of musical possibility, based on a wider variety of algorithms that work with intrinsically musical primitives. This in my view is one of the most important tasks of future algorithmic composers: identify and refine such processes. However, it would be unwise to exclude extrinsic algorithms from the toolkit, as they might expand one's hearing of what is musically acceptable.
\item  It may become feasible to compute a dense parametric map of a more or less compositionally universal algorithm in a reasonable amount of time. In fact, if $\mathsf{P} = \mathsf{NP}$ is ever proved, someone may succeed in creating an algorithm that can compute a dense parametric map of an effectively universal algorithm in \emph{polynomial} time. 

The result might be, again, a God's-eye view of possible structure in music. But I'm not holding my breath, and I think we must continue to work within a garden, it is to be hoped one that grows ever larger and richer, of disparate compositional algorithms.
\end{itemize}


\bibliographystyle{apacite}
\bibliography{gogins_deduplicated}

\end{document}